\documentclass[journal]{IEEEtran}

\usepackage{graphicx}
\usepackage{url}
\usepackage{amssymb}
\usepackage{algorithmic}
\usepackage{algorithm}

\begin{document}
%
\title{Agent-Based Modeling of Systems of Systems}
%
%
%

\author{Jean-Baptiste~Soyez,
	Gildas~Morvan,
	Rochdi~Merzouki
	and~Daniel~Dupont
}

%
%

%

\maketitle

\begin{abstract}
This paper deals with the generic modeling of system of systems (SoS) using agent based modeling (ABM).
SoS are large scale systems including numerous--possibly heterogenous--interacting component systems (CS) 
evolving a dynamic environment.
The goal of this article is to provide a generic formalism which allows to represent and control
the whole complexity of a SoS using agent-based simulations.
In particular organizational aspects of SoS are managed with Agent-Group-Role (AGR) concepts.
Functional aspects, guiding SoS to accomplish their global goals, are handled via a functional specification.
Multi-level aspects are modelled with the Influence Reaction Model for Multi-Level Simulation (IRM4MLS) agent-based meta-model.
Models genererated using this formalism encompass static and dynamic aspects of SoS.
They consider reorganization of SoS caused by changes of goals or sub-system capacity.
All these elements are illustrated in this article using a SoS case study of Intelligent Automated Vehicles (IAV)
initiated by the InTrade european project to automate port container logistic.
\end{abstract}


\IEEEpeerreviewmaketitle

\section{Introduction}

\IEEEPARstart{T}{his} article is concerned with the design and control of artificial complex systems,
i.e., large scale systems composed of numerous communicating components \cite{Aniorte:2006}.
Complex system engineering methods generally take into account the hierarchical organization of such systems.
This is particularly true for systems that are themselves composed of complex systems like federations of systems (FoS).
FoS can be differentiated from conventional systems by their high degree of autonomy,
heterogeneity and spatial dispersion of their components \cite{Sage:2001,Sage:2007}.
Systems of systems (SoS) can be considered as particular type of FoS:
component systems (CS) operate independently and are driven by their local goals,
nonetheless they have to cooperate in order to fufil global goals potentially contradicting local goals \cite{Huhn:2011}.
SoS is a subject of growing interest in many research domains: such as biology, social sciences or military simulations \cite{Bar-Yam:2004}.
Indeed, it is a promising concept that should be able to catch the whole complexity of such real systems.\\

\begin{figure}[!ht]
\centering
\includegraphics[width=1\columnwidth]{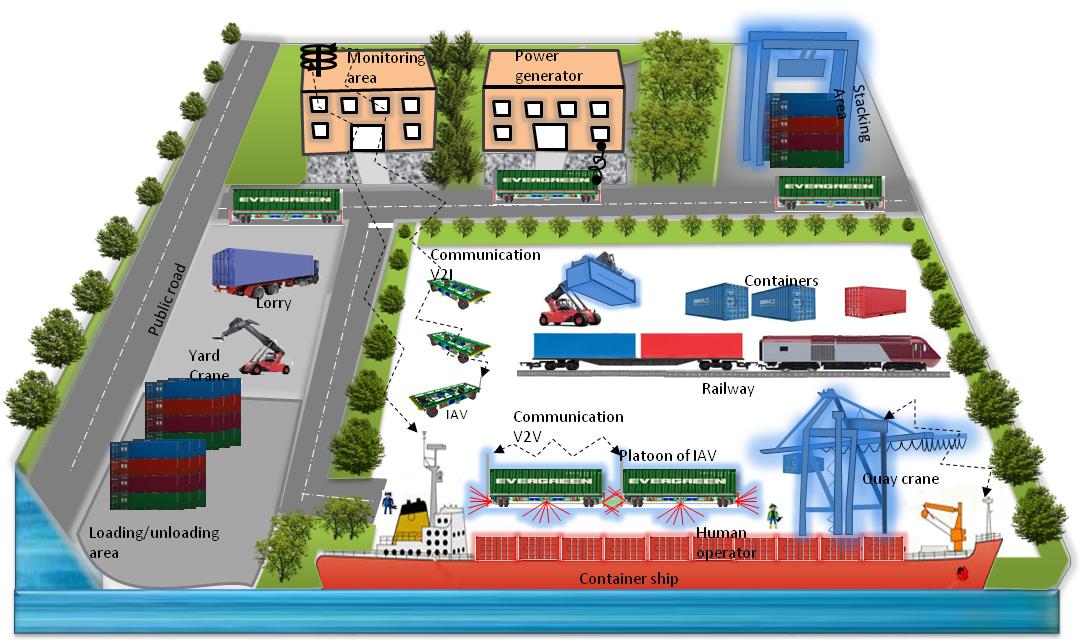}
\caption{Illustration of InTrade SoS complexity.}
\label{intradeComplexity}
\end{figure}

Our work takes place in the frame of the european project InTrade \cite{intrade}.
It deals with automated loading and unloading container operations in large-scale ports of northwest Europe (Dublin, Oostend, Le Havre and Rouen),
using Intelligent Autonomous Vehicles (IAV) controlled by a central operator.
IAVs are semi autonomous vehicles, in charge of container transport tasks.
IAVs can react to changes in environment or goal affectation, they can also reason about their internal dynamic and communicate with human operators or other IAVs.
The container terminal system consists of several IAVs, cranes, operators and boats.
It can be seen as a SoS with numerous heterogeneous elements and several organizations present at different levels and communicating with each other.
To illustrate the proposed SoS modeling and simulating approach, a single practical case,
taken from the InTrade project will be considered throughout this article.\\

\begin{figure}[!ht]
\centering
\includegraphics[width=1\columnwidth]{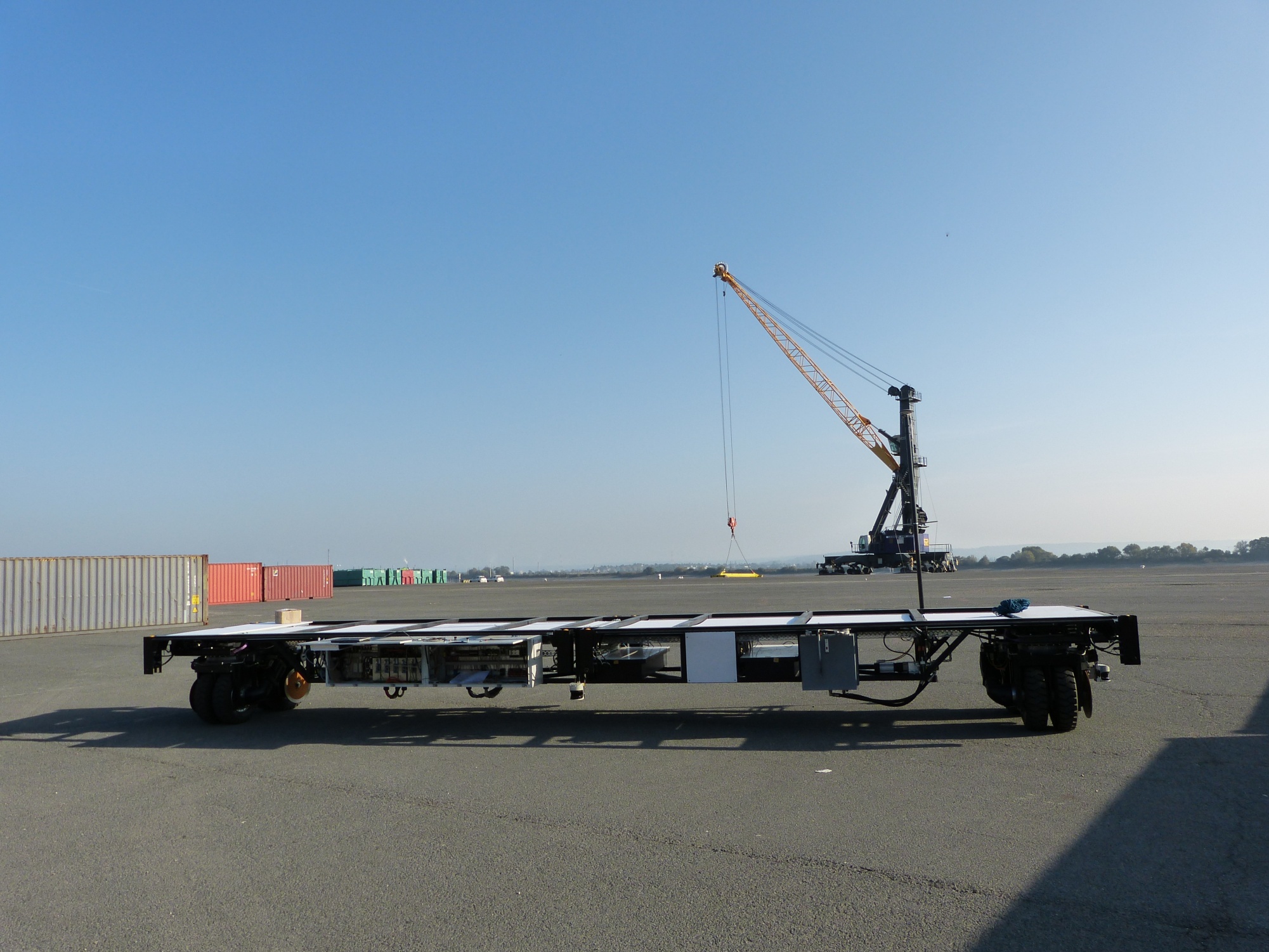}
\caption{Photo of a IAV in a port container.}
\label{IAV}
\end{figure}

To some extent, IAV technology is very similar to Autonomous Guided Vehicle (AGV) technology.
It exists a rich litterature interested on issues generated by the control of numerous AGVs \cite{Le-Anh:2006}.
But, approaches considering IAVs like AGVs with just more intelligence and independance
fail to capture the real complexity of such systems.
A SoS model seems adapted to consider organizational and functional aspects unfolded on several levels of an IAVs fleet.
That is why this article example modeling concerns a IAVs fleet.\\

Our goal is to create a modeling tool for SoS in order to engineer and operate such systems.
To do so, we need a formalism able to define a SoS in terms of behaviour of its CS. 
However, it does not exist any generic, rigorous and complete enough modeling method for SoS.
In this article we introduce a formalism able to capture the dynamic and static aspects of SoS.
This formalism allows to divide the complexity of a system by splitting it into
a set of levels and organizations that represent the system at different scales and structured by groups of CS.\\

\subsection{Related Works}

Because SoS have been introduced in various domains, dedicated tools have been proposed to model and operate them.
Nevertheless, until recently, SoS remained a theoretical concept with no generic simulation formalism.
Several modeling approaches of SoS are non generic and focus on domain-eated issues: \cite{Parker:2010} proposed a classical modeling framework,
defining interfaces between existing CS.
\cite{Huhn:2011} used the agent-based modeling (ABM) paradigm to represent situated CS constrained by their place in a discretized environment (locality).
\cite{Sloane:2007} introduced a model representing the effects of climate policy using also an ABM, nested in a Putman two level game.
\cite{Simpson:2009} focus on the CS connections represented by N square matrix.

\cite{Held:2008} also developed a formalism to represent SoS models, using metrics and attributes.
But this representation focuses on data more than on SoS entities.
The proposed tool is designed to evaluate and predict the performances of a SoS rather than controlling one.

\cite{Khalil:2012a} proposed a formalism based on hypergraphs to represent directed SoS
but he did not provide generic mechanisms to ensure that the constitutive SoS characteristics are preserved during their evolution
(adding or deleting CS, change of general goal).

\cite{Zhou:2011} proposed a generic modeling method for SoS using ABM.
Thus, SoS are endowed with a structure that constraint interactions between agents(communication network or topological environment).

\cite{Gezgin:2012} proposed an ABM approach where CS propose services that define roles into communities.
It focuses on SoS reconfiguration, based on evaluations of the chance to reach a given goal with a good performance.
The reconfiguration process is based on rules adapted to graph grammar.
Models only have two levels: the whole SoS and its CS.
One weakness of this approach is the lack of functional specification for SoS goals
which forces the model to try several configuration before finding the optimal one.
This process works by enhancing the current SoS configuration but there is no guarantee that it will converge.

It seems that the most generic attempts to model SoS rely on ABM. Indeed, as it is shown later,
there are many similarities (and some differences though) between ABM and SoS,
such as the autonomy property of agents and CS.\\

\subsection{Contributions}

The contributions of this article, are of two orders:
1) select and enhance a SoS definition generic and precise ; 2) develop a generic SoS formalism.
The SoS definition is generic enough to be accepted by the SoS community and concrete and precise enough to be operational in a modeling and simulation context.
The proposed formalism allows to easily apprehend the system structure and organizations.
Models generated using this formalism fit to represent IAVs fleet of the InTrade project which are SoS directed by a global goal.\\

Resulting models should contain simple independent CS forming organizations able to accomplish some tasks that single CS cannot accomplish which is possible in MAS.
To make the model easy to apprehend and to divide its computational complexity we will cut the SoS by scales or
independent aspects using the multi-level agent-based modeling metamodel IRM4MLS \cite{Morvan:2010}.

During SoS reconfiguration, caused by global goal or environment change, in \cite{Zhou:2011} and \cite{Gezgin:2012} models,
new SoS configuration are established applying rules contained in CS or in obligations out the system.
There is no independent representation of the global goal to decompose it and establish various plans to accomplish it.
Also there is no independent organizational representation of groups of CS
to choose the best SoS structure to complete previous plans.
Organizational and functional definitions allow to produce such representations.
So, we take inspiration from the $\mathcal{M}oise^+$ model which contains these descriptions and a deontic one which links the previous two
in scope to let the maximum liberty to agents while constraining them to accomplish a global goal \cite{Hubner:2002} \cite{Hubner:2002a}.
$\mathcal{M}oise^+$ is based on the AGR model and reuse its main concepts.\\

This formalism allows the modeling and simulation of SoS in order to control,
diagnose and assess performances of real systems with the same characteristics.
Some approaches highlight the evolutionary aspect of SoS,
but only few
consider SoS in a dynamic way including SoS reconfiguration.
Thus, this article proposes to integrate control mechanisms in SoS to preserve their characteristics during their evolution
like changes in their environment or in the global goal definition.\\

\cite{Held:2008} highlights a characteristic resulting from the 5 SoS fundamental characteristics.
It is the fact that all CS are not needed to fufil the global goal.
That does not mean that SoS have to possess redundant CS to overcome the failure of any CS during its mission accomplishment.
It means that in case of failure, SoS have to be able to reallocate the faulty mission to an available CS or one created for this purpose.
It induces less the obligation of the feasibility of such a reconfiguration than the presence of reconfiguration mechanisms and
the insurance that if the reconfiguration is possible it will be completed.\\

Therefore, according to evolutionary aspect of SoS, a formalism to repsent SoS
should possess two mechanisms: any CS can detect that it is not able to fufil its missions and,
if it serves the global goal, transmits this information to the CS which has authority on it and then can do the same.
The other mechanism concern these failing CS by the reconfiguration of the SoS with the reallocation of missions and/or the organizations change.
These mechanisms and concepts are exposed later in this article.

\subsection{Plan}

This article is organized as follows.
Section II defines what is a System of Systems SoS, 
what are the different types of SoS and gives a graphical representation for SoS.
Section III introduces the concept of Multi-Agent Systems (MAS) and presents ABM tools,
adapted to represent complex systems like SoS.
Then the Multi-level Agent-based Model (MAM) IRM4MLS is introduced.
In section IV, we give static and dynamic elements necessary to describe Multi-Agent SoS (MASoS)
from an organizational point of view and represent its multi-level structure.
After that, we show how to link the previous elements to achieve system-level goals to allow the evolving execution of a MASoS.
Finally, we give some conclusion in section V.

\section{System of Systems (SoS)}

In this section we present the organizational concept of SoS and show some applications.
Then, we explain the need of a solid formalism to represent SoS.
After that, we expose the different families of formalisms apt to represent SoS.
Finally, a graphical notation is introduced  to represent SoS in the rest of this article.

\subsection{SoS Definition}

Many definitions of SoS have been proposed in the literature.
This article leans on Sage and Coppan's definition \cite{Sage:2001,Sauser:2010},
inspired by Maier's  one \cite{Maier:1996}, that can be summarised as follows:\\

\begin{quotation}
A SoS is a set of autonomous Component Systems (CS) endowed with a global goal.
CS can be heterogenous.
They manage theirs own resources and under CS in an independent way and can coexist and cooperate to accomplish a mission that a single CS cannot realize.
CS are geographically distributed without any physical link.
A SoS has to be robust and adaptive: its environment,
goal or structure (by adding/deleting CS) can evolve without modifying its capacity to fufil its global goal.\\

An \textbf{under CS} of a given CS is one of its CS, in a lower level, forming it.
For a given CS, its \textbf{super CS} is the CS, in a higher level, to which it belongs.\\
\end{quotation}

This definition respects the five following characteristics illustrated with an example from the InTrade project.

\begin{enumerate}

\item{\textbf{Operational independence}}:  Each CS possesses its own resources necessary to accomplish its mission.
For example, an IAV has its own stock of energy and its personal status diagnosed online.

\item{\textbf{Managerial independence}}: Once a mission is attributed to a CS, it manages itself and its under CS, on its own, to accomplish it.
For example, a fleet of IAVs attached to a quay decides on its own how to organize its IAVs to accomplish its allocated missions.

\item{\textbf{Geographic distribution}}: CS can exchange information but there is no physical exchange of energy.
For example, when two IAVs are attached to form a platoon to be able to carry 40 feet containers, they lost their geographic distribution and cease to be considered as CS.

\item{\textbf{Emergent behaviour}}: The global goal of a SoS can only be reached by the joint action of more than one CS.
For example, if one IAV is sufficient to reach the SoS goal to perform all transport tasks in the port, the SoS has no reason to exist. 

\item{\textbf{Evolutionary development}}: A SoS can adapt itself to the addition or deletion of CS or
evolution of the environment or change of  the global goal to stay able to reach it by the actions of CS.
For example, when an IAV endowed with a mission lost its capacity to fufil it, it informs its super CS.
This super CS has means to overtake this loss of capacity by allocating the mission to another CS agent with intact capacities.\\

\end{enumerate}

Similar characterizations like Bordman and Sauser's one \cite{Boardman:2006} exist.
This characterization provides the following characteristics differentiating an SoS from monolithic systems:
1) \emph{autonomy} - CS exercise autonomy in order to fufil the SoS global goal;
2) \emph{belonging} - CS choose to belong to the SoS based on cost/benefit basis;
3) \emph{connectivity} - CS provide dynamic connectivity to enhance overall SoS capability;
4) \emph{diversity} - a product characteristic of an SoS not avaible from single systems;
5) \emph{emergence} - capability is provided that was not originally foreseen during development,
leading to early detection and elimination of undesirable behaviors.
\cite{Gorod:2008} establishes a review of SoS engineering (SoSE) methods and
uses previous SoS characterization to create a SoSE management framework.\\

\cite{Jamshidi:2008} gave a SoS definition recognized in the SoS community:
\begin{quotation}
``Systems of systems are large-scale integrated systems
that are heterogenous and independently operable on their own,
but are networked together for a common goal''.\\
\end{quotation}

Other definitions seem to be too vague to be exploited here.
In many cases \cite{Maier:1996,Dauby:2011} they only describe information systems where
interactions between systems are limited to information exchanges.
This is too restrictive to represent systems in strong interaction \cite{Michel:2003}, e.g.,
two robots joining each other to carry an object too heavy for only one robot.
A more comprehensive definition, in relation with the MAS paradigm, is given in section IV.\\  

\subsection{Types of SoS}

The SoS concept has many applications in different domains such as military and maratime simulation, biology, climatology and sociology
\cite{Bar-Yam:2004, Gerst:2012, sdsFAQ:2004, Mahulkar:2009}.
In sociology, the gregarious behavior of humans can lead to naturally emergent organizations or societies, considered as SoS.
In a military context, SoS are sometimes the only available approach to represent influences between organizations at different levels,
with their own area of intervention  (land, air, sea) in large-scale operations.\\

In its guide to engineer SoS, the US Departement of Defense (DoD) describes four types of SoS,
depending on their degree of centralization \cite{SoSDoD:2010}.
They can be: 1) \textbf{Virtual}. The SoS lacks a central management and a centrally agreed-upon purpose.
2) \textbf{Collaborative}. CS within the SoS interact more or less voluntary to fufil agreed-upon central purposes.
3) \textbf{Acknowledged}. The SoS has acknowledged objectives, a designated manager, and resources,
while CS retain their independent ownership, objectives, funding, development, and sustainment approaches.
4) \textbf{Directed}. The SoS is built to fufil specific purposes. CS operate independently,
but their operational mode is subordinated to central management purposes.\\

Virtual SoS are generally used to study the evolution of natural complex systems
such as social or biological systems \cite{Bar-Yam:2004}.
And they put forth the hypothesis that purely collaborative SoS emerge only if the environment influences the CS,
posing a new problem which requires the SoS existence.
These cases do not match the subject of this article, which focuses on engineering systems endowed with a hierarchy and nested level structure.
Therefore, this article deals mainly with directed or acknowledged SoS.

\subsection{Graphical Representation and Particular CS}

In this section a graphical representation of the real entities of a SoS is introduced.
For more details on CS notation see section IV.
It permits to easily see how a CS is constituted by its lower-level CS and distinguishes between inclusion and communication links.
Let consider an example from the Intrade project. A whole fleet of IAVs, $CS_{1,3}$, is composed of quay fleet from quay 1, $CS_{1,2}$, and quay 2, $CS_{2,2}$.
The $CS_{1,2}$ fleet is composed of a $CS_{1,0}$ platoon (including $CS_{1,0}$ and $CS_{2,0}$ IAVs) and an IAV $CS_{3,0}$.
$CS_{2,2}$ fleet is composed of three IAVs: $CS_{4,0}$, $CS_{5,0}$ and $CS_{6,0}$.\\

\begin{figure}[!ht]
\centering
\includegraphics[width=1\columnwidth]{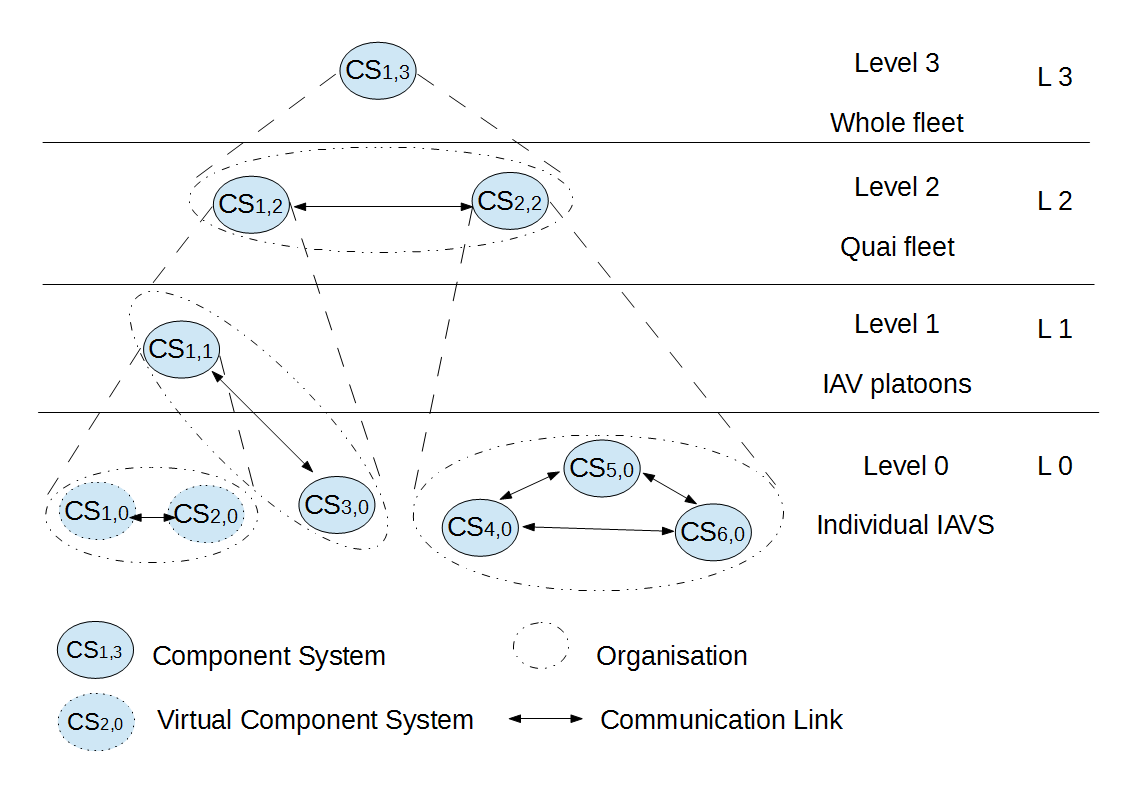}
\caption{Multi-level graphical representation of a SoS.}
\label{graphicalRepresentation}
\end{figure}

Some CS or set of CS can temporarily violate some of the SoS characteristics explained further.
That does not mean the end of the SoS, but that CS may have to be deleted, at least temporarily,
to make the SoS coherent with these characteristics. We call this entities \textbf{virtual component system}.
For example $CS_{1,0}$ and $CS_{2,0}$ are IAVs physically linked to form a platoon.
They partially have lost their independence but still exist.
In addition a CS that cannot be divided into CS respecting these characteristics is called an \textbf{elementary component system}
(like $CS_{1,1}$, $CS_{3,0}$, $CS_{4,0}$, $CS_{5,0}$ and $CS_{6,0}$).
Basically the physical individual entities of a system are considered as elementary CS
(like $CS_{3,0}$, $CS_{4,0}$, $CS_{5,0}$, $CS_{6,0}$ and potentially $CS_{1,0}$ and $CS_{2,0}$).
Opposed to elementary CS, the only CS present at the highest level, $CS_{1,3}$,
which is equivalent to the SoS can be named the \textbf{top component system}.\\

In the following sections of the article we will model CS using CS agents taking advantages of autonomy and independence inherent to agents.
CS agents are agents representing CS in SoS and showing the same characteristics as CS.

\section{Multi-Agent Systems (MAS)}

This section presents MAS modeling tools and how they are applied to represent complex systems like SoS.
In particular the IRM4MLS multi-level metamodel, which allow to represent complex systems at different granularity levels
will be introduced.

\subsection{MAS Definition}

Most authors generally agree to define a MAS as a system composed of communicating and collaborating agents,
that have objectives (personal or collective) and resources to achieve them.
Communication implies the existence of a shared space to support it. This space is generally called environment.
Ferber, in his book \cite{Ferber:1995}, defines agents and MAS as follows.\\

We call agent a physical or virtual entity
\begin{enumerate}
\item which is able to act in an environment,
\item which can communicate (directly) with other agents,
\item which is driven by a set of tendencies (under the form of individual objectives or
a satisfaction or survival function that it tries to optimize),
\item which possesses its own resources,
\item which is able to perceive (in a limited way) its environment,
\item which disposes only of a partial representation (eventually none) of this environment,
\item which possesses competencies and offers services,
\item which can eventually reproduce itself,
\item  whose behaviour tends to satisfy its objectives with its resources and its competencies and according to its perceptions, representations and communications it receives).\\
\end{enumerate}

We call Multi-agent System (or MAS), a system composed by following elements:
\begin{enumerate}
\item An environment $E$, i.e, a topological space, that generally embodies a metric,
\item A set a objects $O$. These objects are situated: at a given time,
it is possible to associate to each object a position in $E$.
Objects can be perceived, created and modified by agents,
\item A set of agents $A$, which are particular objects ($A \subseteq O$), that represent the active entities of the system,
\item A set of relations $R$ that link objects (and then agents) between them,
\item A set of operators $Op$ allowing agents of $A$ to perceive, produce, consume, transform and manipulate objects of $O$,
\item Some operators dealing with the representation of the application of previous operators and
the world reactions of modifications, which can be called universe laws.
\end{enumerate}

\subsection{Agent-Based Modeling of Complex Systems}

MAS are widely used to simulate interactions between complex, autonomous entities acting in parallel.
This kind of entities are easily found in complex systems like SoS.

\cite{Scerri:2010} presents a MAS model to represent systems with the High Level Architecture (HLA).
HLA is a metamodel that allows to make coexist and coordinate different simulations in the same platform.
Even if HLA is more concerned with implementation than our approach,
the proposed model deals with coexisting independent entities forming different levels of organization.

\cite{Picault:2011} also presents a MAS model, PADAWAN, whose specificity is to include agents and their nested environment in other agents.
This way of seeing a system with aggregate organization at a level creating individuals at a higher level is common with SoS.

\cite{Morvan:2012} proposes a bibliography to list the existing multi-level agent-based modeling approaches.
This article identifies the different theoretical issues, meta-models, platforms and applications related to that domain.\\

First of all a SoS should be considered as an organizational concept,
and therefore it can be compared to MAS organizational models such as AGR (Agent, Group, Role) \cite{Ferber:1998,Ferber:1999}.
This formalism organizes a system into groups depending in which agents play roles.

\cite{Gaud:2008} uses a hierarchical multi-agent metamodel (CRIO) based on the concept of holon~\cite{Koestler:1967}.
A holon represents both an individual entity and a set of organizations composed with entities with the structure.
CRIO contains also an organizational decomposition aspect.
In his thesis he presents a complete methodology to model and implement a complex system. He used ASPECS,
an engineering software process which describes all necessary steps to conceive a software. It is based on CRIO.\\

\cite{Hubner:2002, Hubner:2002a} studied organization oriented MAS and to insure that the system will complete its global goal
while guaranteeing agents autonomy in terms of behaviour and creation of organizations,
they use deontic specification to link groups and allocation of missions in their model: $\mathcal{M}oise^+$.
In a way this is similar to our approach to SoS.
But in our approach, the existence of groups is strongly constrained by the necessity to fufil the global goal of the system.

\cite{Ribino:2012} created a metamodel mixing holonic organizational aspects of ASPECS and deontic specification
(renamed norms) issued from $\mathcal{M}oise^+$ to instantiate holonic MAS (HMAS).
The created HMAS are designed to reach a global goal with coordinating independent entities respecting the system integrity during its changes.
We can inspire ourselves of this approach if we consider the common points and the difference between SoS and holonic systems.

\subsection{Influence Reaction Model for Multi-Level Simulation (IRM4MLS)}

A SoS is a specific case of complex system whose elements are interacting at different levels.
Such multi-level aspects of SoS can be represented using IRM4MLS.
Another interest to represent a SoS with a multi-level model is to divide the complexity of the system
and make it easier to apprehend and compute by separating scales of organizations  or independent aspects of a SoS.\\

IRM4MLS is a MAM.
This meta-model allows to represent multiple entities in interaction situated in different levels.
Here a level is considered in a general way and not obligatory as a scale.
Moreover interactions between entities in different levels and their results are represented without any bias
thanks to the influence reaction model used in IRM4MLS \cite{Ferber:1996, Michel:2007}.\\

A IRM4MLS model is characterized by a set of levels, $L$, and relations between levels.
Two types of relations are considered in IRM4MLS: influence (agents in a level $l$ are able to produce influences in a level $l' \neq l$) and
perception (agents in a level $l$ are able to perceive the state of a level $l' \neq l$).
These relations are respectively formalized by two digraphs,
$<L,E_I>$ and $< L,E_P>$ where $E_I$ and $E_P$ are sets of edges, i.e.,
ordered pairs of elements of $L$.
The dynamic set of agents at time $t$ is denoted $A(t)$.
$\forall l \in L$, the set of agents in $l$ at $t$ is $A_l(t) \in A(t)$.

Let $\delta(t) \in \Delta$  be the dynamic state of the system at time $t$: $\delta(t)=<\sigma(t),\gamma(t)>$, where $\sigma(t) \in \Sigma$
is the set of environmental characteristics and $\gamma(t) \in \Gamma$ the set of influences representing system dynamics.
The state of a level $l$ at time $t$ is noted: $\delta_l(t)=<\sigma_l(t),\gamma_l(t)>$, where $\sigma_l(t) \in \Sigma_l$ and $\gamma_l(t) \in \Gamma_l$
are respectively set of environmental characteristics and influences representing dynamics, both in $l$.

\begin{figure}[!ht]
\centering
\includegraphics[width=1\columnwidth]{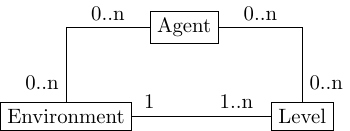}
\caption{Central concepts of IRM4MLS (cardinalities are specified with the UML way)}
\label{IRM4MLS}
\end{figure}

The state of an agent $a \in A$ is characterized by :

-its physical state $\phi_a \in \Phi_a$ with $\Phi_a \subset \Sigma_a$ (e.g, its position)

-its internal state $s_a \in S_a$ (e.g., its beliefs)\\

An agent acts in a level if a subset of its physical state belongs the state of this level.
An agent can act in multiple levels at the same time. Environment is also a top-class abstraction.
It can be viewed as an agent with no internal state that produces influences,
representing the natural dynamic of environmental properties, in the level.
The scheduling of each level is independent: models with different temporalities can be simulated without temporal bias.
On the another hand, it permits to execute only the relevant processes during a time-step.
A major application of IRM4MLS is to allow microscopic agents (members) to aggregate and form-up lower granularity agents (organizations).
It can be useful to create multiple levels at a same scale to represent different domain parts of the same phenomenon.
In the rest of this article we consider that two levels are at the same scale if they have the same spatial and temporal extents.\\

If levels form a hierarchy of nested scales, they can be figured by a hierarchical graph presented in \cite{Soyez:2012},
which indicates the level imbrication.
A simple edge represents an inclusion link between two levels.
For example, an ($l_1$,$l_2)$ edge signifies that $l_2$ has higher spatial or temporal extents than $l_1$.
Then the agents situated in $l_1$ can be aggregated and the resulting aggregate can be instantiated in $l_2$.
A pair of symmetric edges means there is a complementarity link between two levels.
For example, the $(l_1,l_3)$ and $(l_3,l_1)$ pair of edges mean that $l_1$ and $l_3$ are at the same scale.
Thus agents simultaneously present and activated in $l_1$ and $l_3$ can figure different aspects of the same entities.

\begin{figure}[!ht]
\centering
\includegraphics[width=1\columnwidth]{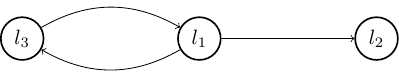}
\caption{Example of hierarchical graph.}
\label{hierarchicalGraph}
\end{figure}

\section{Agent Formalism to represent SoS}

MAS definition lacks concepts such as an organizational and a hierarchical definition to define which CS integrates and/or allocates missions to another CS,
to be able to represent a SoS.
Also, the notion of global goal and missions are expressed without taking in account SoS multi-level structure.
This makes harder the SoS to reason about them to reorganize itself in order to fulfil them.
These concepts are crucial because they are needed to express the characteristics of managerial independence, evolutionary development, cooperation and coexistence.\\

In this section, we give all static (1) and dynamic (2)  elements necessary to represent a SoS and that are not basically included in all MAS:
1) an organizational description of entities and  a multi-level environment and 2) two mechanisms to conduct SoS initialization and capacity evolution.
Then, we give some elements to prove that multi-agent models, created using our formalism, respect SoS fundamental characteristics.

\subsection{Static Apects}

SoS is an organizational concept in which CS are entities that can  be individuals or aggregations of CS in lower level(s).
A good way to represent these groups and provide the super CS a way to reason about the mission distribution
to reach its allocated goal is to use the Agent-Group-Role model (AGR) \cite{Gutknecht:2001, Ferber:2004}.\\

From the AGR point of view, an agent is an entity playing a set of roles in several groups.
An agent is characterized by its roles, knowledge and capacities.
A \emph{capacity} is the abstract description of a knowledge.
It regroups the means that allow to accomplish a task.
This represents a logical functionality for providers (owners) or  applicant (users) entities.
In a more prosaic way, an agent or group capacity is the possibility for this entity to accomplish a goal,
i.e., reaching a target level state, by the means of its actions, in a precise given level state.\\

\begin{figure}[!ht]
\centering
\includegraphics[width=1\columnwidth]{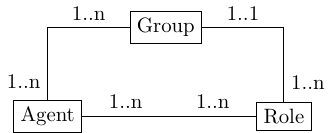}
\caption{Central concepts of AGR (cardinalities are specified with the UML way)}
\label{agrConcepts}
\end{figure} 

An \emph{agent role} is a concrete instance of a role.
It describes the behaviour in context defined by a group.
It confers a status and means in this group to interact with other agents playing roles in that group.
Before the initialisation of the SoS, groups are just definitions and a same set of agents can instantiate several groups sequentialy or instantaneousy.
These groups possess capacities determined by the roles played by agents and their individual capacities.\\
 in function of roles played by its agents and their capacities.\\

\subsubsection{CS Modeling}

To designate in a unique way and without any ambiguity a CS in a SoS we can use the following formalism:
$CS_{n,l}$ designates the n$^{th}$ CS  and its representing agent in the level $l$. Here is a definition inspired by Gaud.
A CS, $CS_n$ of a level $l$, can be noted as follows:\\

$CS_{n,l}=<CS_{n',l+1},R_{n,l},\mathcal{CS}_{n'',l-1},OP,\Psi,\Lambda>$, with :

\begin{itemize}
\item{$CS_{n',l+1}$}: a super CS which is directly composed of $CS_{n,l}$ and possibly other under CS. This CS has to be in a level higher than $l$.

\item{$R_{n,l}$}: the set of roles played by $CS_{n,l}$, $R_{n,l}=2^{Roles(O_{n,l})}$ with $O_{n,l}$ the set of groups in which $CS_{n,l}$ plays at least one role.

\item{$\mathcal{CS}_{n'',l-1}$}: the set of under CS constituting the super CS: $CS_{n,l}$. These CS have to be in level(s) lower than  $l$.

\item{$OP$}: the set of groups taking part in life and functioning of $CS_{n,l}$ CS and contributing to accomplish the objectives linked to the $R_{n,l}$ roles.

\item{$\Psi: \mathcal{CS}_{n'',l-1} \to 2^{Roles(OP)}$}: A function associating an under CS member to the set of roles that
it plays in groups defined in $CS_{n,l}$, such as $\forall csi \in \mathcal{CS}_{n'',l-1}, \Psi(csi) \neq  \emptyset$.
The role function provides the set of all roles defined in $OP$ groups.

\item{$\Lambda: \Sigma_l*\Sigma_l \to \{0,1\}$}: A capacity function indicating if a goal state in $l$ is reachable, by the action of $CS_{n,l}$, knowing the actual state of $l$.  This function depends on the evolving capacities of $CS_{n,l}$ and also on the capacities offered by $OP$ to $CS_{n,l}$ when $CS_{n,l}$ is not an elementary CS.\\
\end{itemize}

In the case of SoS engineering there is a hierarchy of levels of increasing abstraction, approximation and scale.
These levels can be represented by a hierarchical graph, as shown above, which indicates the level imbrication.
When a level $l_i$ is directly included in a level $l_j$, CS in $l_i$ compose higher scale CS in $l_j$.
A CS can directly belong to only one CS of a higher level, therefore, here, a level can be directly included in at most one level.
When a CS $CS_{i',j'}$ is an under CS of $CS_{i,j}$ it is noted $CS_{i',j'} \in CS_{i,j}$\\

\subsubsection{Group of CS Modeling}

To decide how to form super CS from organization of under CS and how CS are organized to reach goals it is necessary to endows our model with set of group specification defined as follows:\\

$gs=_{def}<R,L_R,\mathcal{L},C^{intra},C^{inter},nr,nc>$, with

\begin{itemize}
\item{$R$}: the set of non abstract roles playable by agents in groups created from $gs$.

\item{$L_R:R_{gs} \to L$}: a function  indicating the level of an agent which can play a given role in $gs$.

\item{$\mathcal{L}$}: indicates the level(s) of the group created from $gs$.
if $\mathcal{L}$ contains only one level $l$ and $l$ is higher than any level of $L_R(R)$ then
the group defined by $gs$ has to be instanciated by a CS agent situated in $l$.
CS agents playing role(s) in $R$ are the under CS of this new CS agent.

\item{$C^{intra}$}: a compatibility function which indicates if two roles are compatible in the same group.
By default, two roles are not compatible. $\rho_a \Delta \rho_b$ says that agents playing role $rho_a$  are authorized to play role $rho_b$. This relation is reflexive and transitive.

\item{$C^{inter}$}: a compatibility function which indicates if two roles are compatible in two different groups. It possesses the same formalism than $C^{intra}$.

\item{$nr:R_{gs} \to N*N$}: a function which specifies the number (minimum, maximum) of roles which has to be played in a group, ex., $nr_{gs}(carrier)=(1,3)$ means that groups emanating from $gs$ have to possess at least one and at most 3 agents playing the $carrier$ role.

\item{$nc:Rgs*C \to N*N$}: a function which specifies for a given role and capacity the quantity
(minimum, maximum) of this capacity which has to be available for each agent playing this role and for the whole population of agents playing this role.\\
\end{itemize}

Here a capacity should be expressed in a functional way and has to be quantified, i.e.,
not expressed in terms of goal states but in terms of concrete results of agent actions.

$nc_{gs}(carrier,carry One Container/hour)=(1,\infty,3,\infty)$ means that an agent playing
the role carrier in a group emanating from $gs$ has to possess at least one unit of the $“Carry One Container/hour”$
capacity and the whole population of agents playing this role in the group has to possess
at least three cumulative units of the $“Carry One Container/hour”$ capacity with in both cases no maximum limit for this quantified capacity.\\

\subsubsection{Multi-level environment and goal decomposition}

IRM4MLS provides a support to define global goals. A global goal is a world state to be reached by the system,
which can be expressed as a description of the system and its environment variables.
A global goal can be divided into a set of missions. This cut out can be functional,
geographic or other. Because the modeling of SoS include multi-level representation, this cut out can be done according to levels.
Thus missions of a level can be expressed as states of this level, i.e.,
the states of CS and environment of this level.
Moreover the global goal of a SoS can be expressed as a goal state of the highest level $L_H$.
It is used to define the global goal of the SoS : $Gg=L_H(t')=<L_H(t'),L_H(t')>$.
$t'$  can be considered as a time limit for the SoS to reach $Gg$.\\

Because it is the most abstract and approximate,
the wished state $L_H(t')$ of the highest level $L_H$ can be divided into a set of states in
every lower level corresponding to local missions which describe more in details the accomplishment of the global goal.
For every inclusion link, $(l_i,l_j)$ with $l_i,l_j \in L$, in the SoS hierarchical graph, there is a function $F_{i,j}$,
which indicates how a goal state in $l_j$ can be translated into several goal states in $l_i$.
Each of these translated goals can correspond to a given mission allocated to CS in $l_i$ that allows to reach the goal of $l_j$.
Because for CS in a given level, there are several ways to fulfil a goal expressed in a higher level.\\

\subsection{Illustration of Static Aspects}

To illustrate the static aspects we model the SoS presented in section II with our formalism.\\

$L=\{l_o, l_1, l_2, l_3\}$ with
\begin{itemize}
\item $l_3 = l_H = levelOfWholeFleet$,\\
\item $l_2 = levelOfQuayFleet$,\\
\item $l_1 = levelOfIAVPlatoon$ and\\
\item $l_0 = levelOfIndividualsIAVs$.\\ 
\end{itemize}

\begin{figure}[!ht]
\centering
\includegraphics[width=1\columnwidth]{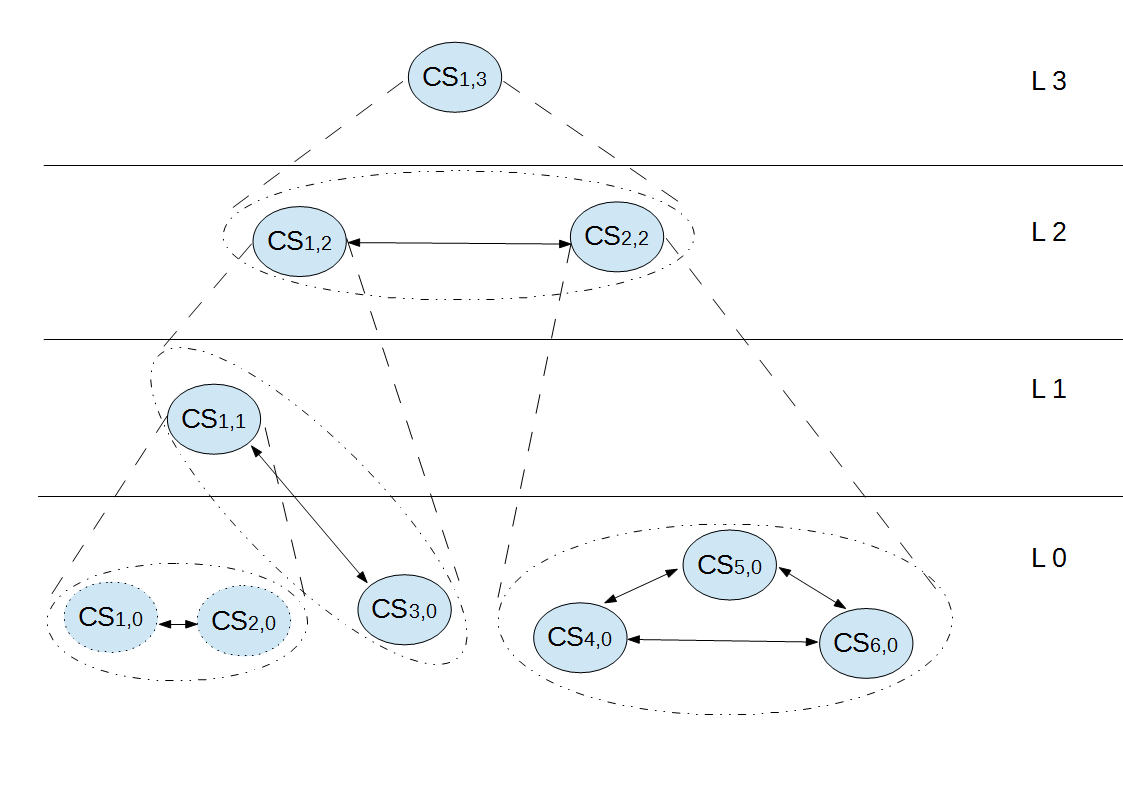}
\caption{Example of a SoS.}
\label{graphicalRepresentationSimple}
\end{figure}

All CS are represented by CS agents.
$\mathcal{A} = \{CS_{1,3},CS_{1,2},$\\$CS_{2,2},CS_{1,1},CS_{1,0},CS_{2,0},CS_{3,0},CS_{4,0},
CS_{5,0},CS_{6,0}\}$\\

Here we only represent in detail $CS_{2,2}$ and its under CS $CS_{4,0}$, $CS_{5,0}$ and $CS_{6,0}$.

\begin{eqnarray*}
CS_{2,2} =  \langle CS_{1,3}, \{quayTransporter, quayTransmitter\},\\
\{ CS_{4,0},CS_{5,0},CS_{6,0} \}, \{g_1\}, \psi_{2,2}, \lambda_{2,2}\} \rangle\\
\end{eqnarray*}

\begin{figure}[!ht]
\centering
\includegraphics[width=1\columnwidth]{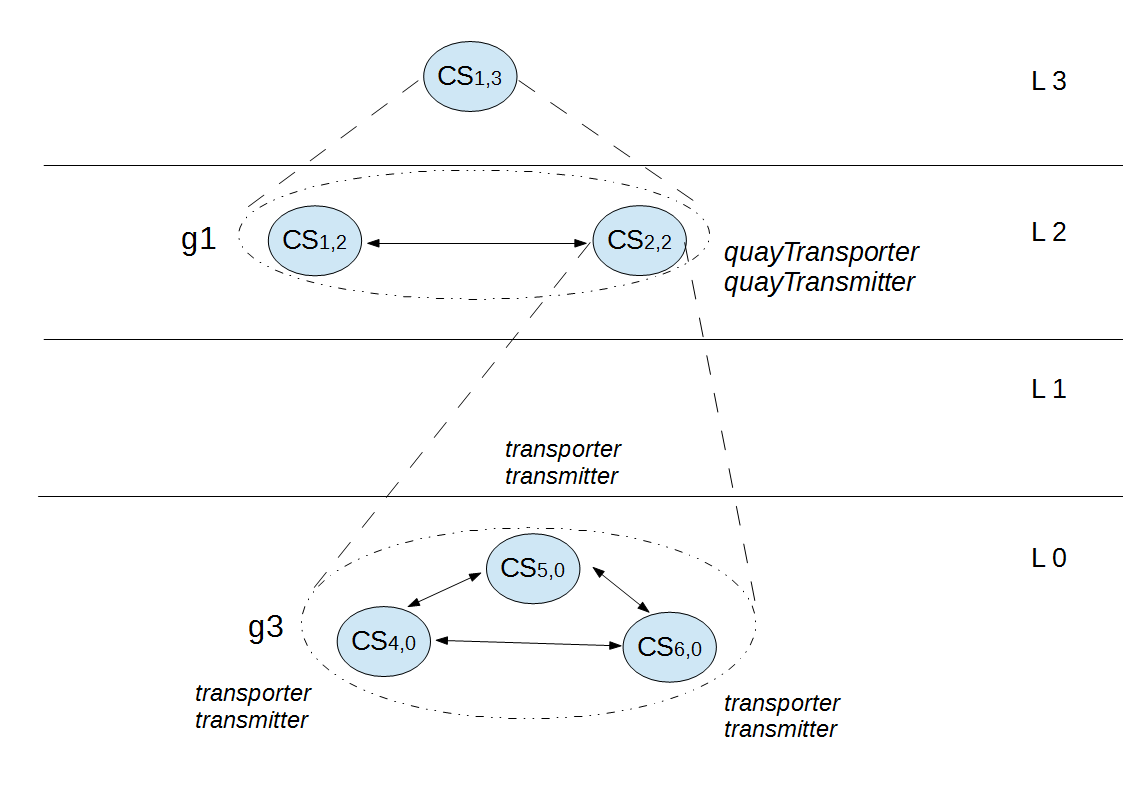}
\caption{$CS_{2,2}$, its under CS and their roles in their group.}
\label{graphicalRepresentationSimple}
\end{figure}

$CS_{2,2}$ is an under CS of $CS_{1,3}$, the whole fleet of IAVs in the port.
It plays the roles $quayTransporter$ and $quayTransmitter$ which consist in managing communication and
container transport of a fleet of individual IAVs attached to a quay.
Its under CS are individual IAVs, $CS_{4,0}$, $CS_{5,0}$ and $CS_{6,0}$.
Here $CS_{1,3}$ is constituted of only one group $g_1$ built of $CS_{2,2}$ and $CS_{2,2}$.
$\psi_{2,2}$ indicates the roles played by $CS_{4,0}$, $CS_{5,0}$ and $CS_{6,0}$ in $CS_{2,2}$.
$\lambda_{2,2}$ is the function indicating if $CS_{2,2}$ can reach its goals in its $l_2$ level.
This function takes in account $CS_{2,2}$ capacities and the individual capacities of its IAVs, $CS_{4,0}$, $CS_{5,0}$ and $CS_{6,0}$

The group specification of $g_1$ can be :

\begin{eqnarray*}
gs_1 =  \langle \{quayTransporter, quayTransmitter\},\\
\{l_{quayTransporter} = \{l_2\}, l_{quayTransmitter} = \{l_2\}\},
\{l_3\},\\
\emptyset,
\{quayTransporter \Delta quayTransmitter\},\\
\{nr_{quayTransporter} = (1,\infty),nr_{quayTransmitter} = (1,\infty)\},\\
\{nc_{quayTransporter} = (1cont/hour,\infty,1cont/hour,\infty),\\nc_{quayTransmitter} = (2mess/sec,\infty,2mess/sec,\infty)\},
\rangle\\
\end{eqnarray*}

CS agents can play the roles $quayTransporter$ and $quayTransmitter$ in groups defined by $gs_1$.
These CS agents are in level $l_2$. These groups produce CS agents in level $l_3$.
When a CS agent plays a role in a group defined by $gs_1$, it cannot play a role in any other group.
A CS agent can play simultaneously both $quayTransporter$ and $quayTransmitter$ roles in a group defined by $gs_1$.
A group defined by $gs_1$ must contain--at least--one CS agent playing the role $quayTransporter$ and, at least, one CS agent playing the role $quayTransmitter$.
A CS agent playing $quayTransporter$ must possess the capacity to transport, at least, one container per hour.
A CS agent playing $quayTransmitter$ must possess the capacity to transmit, at least, two messages per second.
And the whole population of agents playing $quayTransmitter$ must possess the cumulative capacity to transmit, at least, two messages per second.

\begin{eqnarray*}
CS_{4,0} =  \langle CS_{2,2}, \{ transporter, transmitter \}, \emptyset, \{g_3\}, \emptyset, \lambda_{4,0}  \rangle\\
CS_{5,0} =  \langle CS_{2,2}, \{ transporter, transmitter \}, \emptyset, \{g_3\}, \emptyset, \lambda_{5,0}  \rangle\\
CS_{6,0} =  \langle CS_{2,2}, \{ transporter, transmitter \}, \emptyset, \{g_3\}, \emptyset, \lambda_{6,0}  \rangle\\
\end{eqnarray*}

$CS_{4,0}$ is an under CS of $CS_{2,2}$, the quay fleet of quay 2.
It plays the roles $transporter$ and $transmitter$ which consist in transporting a container and communicate with distant IAVs or
the central authority of the port.
$CS_{4,0}$ has no under CS.
Here $CS_{1,3}$ is constituted of only one group $g_3$ built the three individual IAVs, $CS_{4,0}$, $CS_{5,0}$ and $CS_{6,0}$.
$CS_{4,0}$ has no under CS roles to define.
$\lambda_{4,0}$ is the function indicating if $CS_{4,0}$ can reach its goals.
Because $CS_{4,0}$ is a elementary CS, this function takes only into account
capacities of the individual IAVs instanciated by $CS_{4,0}$.
$CS_{5,0}$ and $CS_{6,0}$ are defined in the same way than $CS_{4,0}$.

The group specification of $g_3$ can be :

\begin{eqnarray*}
gs_3 =  \langle \{transporter, transmitter\},\\
\{l_{transporter} = \{l_0\}, l_{transmitter} = \{l_0\}\},
\{l_2\},\\
\emptyset,
\{transporter \Delta transmitter\},\\
\{nr_{transporter} = (1,\infty),nr_{transmitter} = (1,\infty)\},\\
\{nc_{transporter} = (1cont/hour,\infty,1cont/hour,\infty),\\nc_{transmitter} = (2mess/sec,\infty,2mess/sec,\infty)\},
\rangle\\
\end{eqnarray*}

$gs_3$ is a group specification very similar to $gs_1$ with a change of level and scale.


\subsection{Dynamic Aspects}

In this section we present the mechanisms that insure the feasibility of the global goal during SoS evolution.
The first one concerns the creation of groups with capacity matching accomplishment of the global goal and the second concerns the change of capacities for CS.\\

There are several articles deal with SoS construction or reconfiguration \cite{Acheson:2012,Yang:2011},
some including the management of CS capacity \cite{Adler:2012,Flanigan:2012},
but without addressing the control of such systems.\\

\subsubsection{SoS Creation}


In our approach we try to incite CS agents to adopt a organizational structure appropriated to cooperate and fufil a global goal.\\

At the SoS initialisation, outside the top CS and all CS with a physical existence, like most of the elementary CS,
all intermediary CS have to be instanciated.
The situation is quite similar when the global goal or even a non elementary CS goal is modified:
all its under CS except those with no under CS (virtual or not) have to be created again.
This is done in order to design organizations with capacities adapted to reach the new goal and its sub-goals.
To form organizations of CS matching the SoS global goals needs our model
has to lean on a functional specification.\\

Missions can be seen as sequences of CS agent actions required to meet a goal. 
Missions in that model express the fact that a CS agent can engage simultaneously to fufil several goal with the same mission.
The fact that a mission $\mathcal{M}x$ is allocated to the $CS_{n,l}$ CS is noted $\mathcal{M}x_{n,l}$.\\

The global goal is decomposed as a graph tree. Each time a goal is divided into subgoals with a change of scale,
it shows that these goals will be accomplished by a CS in one level and its under CS in at least another level.
A goal $g_x$ can be noted with its corresponding level $l$: $g_{x,l}$.\\

Here follows a graph tree presenting the possible global goal allocated to the SoS presented in fig. 1.
CS agent $CS_{1,3}$ would be engaged on $\mathcal{M}1$, $CS_{1,2}$ on $\mathcal{M}2$, $CS_{2,2}$ on $\mathcal{M}3$ and so on.
So the set of missions for this goal decomposition should be equal to
$\mathcal{M} = \{ \mathcal{M}1_{1,3}, \mathcal{M}2_{1,2}, \mathcal{M}3_{2,2}, \dots \}$\\

\begin{figure}[!ht]
\centering
\includegraphics[width=1\columnwidth]{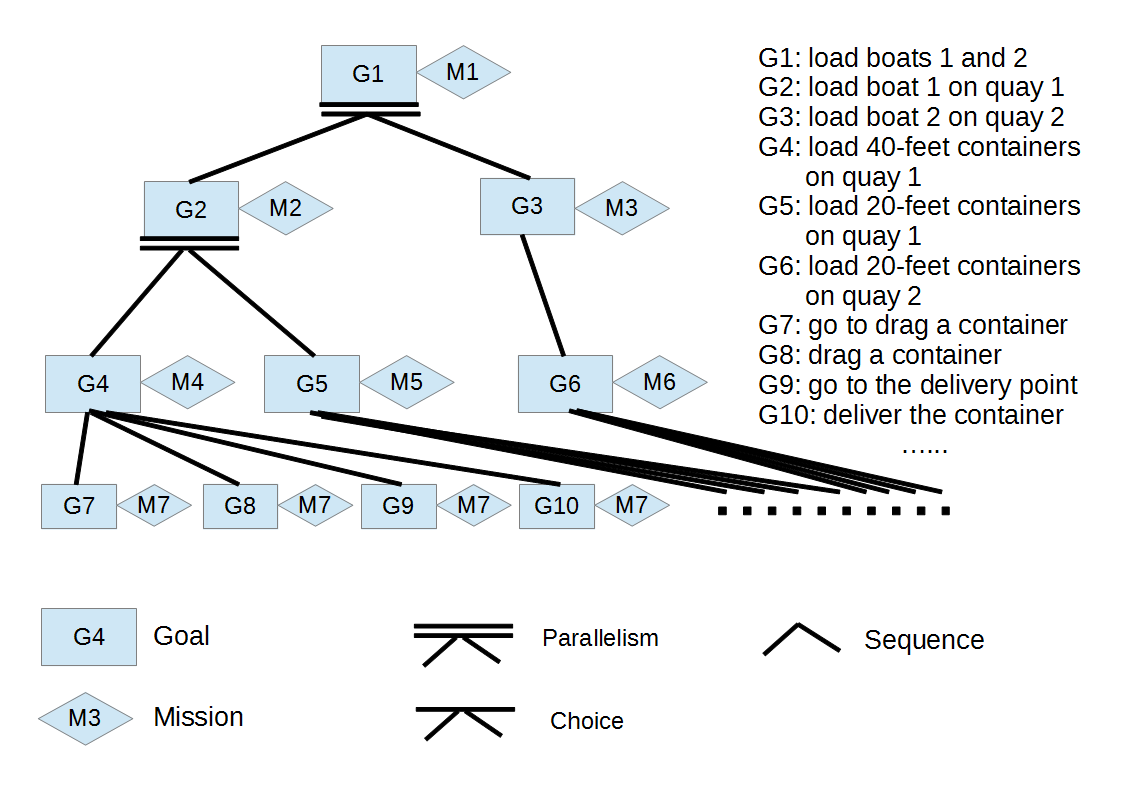}
\caption{The functional decomposition of SoS global goal.}
\label{goalDecomposition}
\end{figure}

\begin{figure}[!ht]
\centering
\includegraphics[width=1\columnwidth]{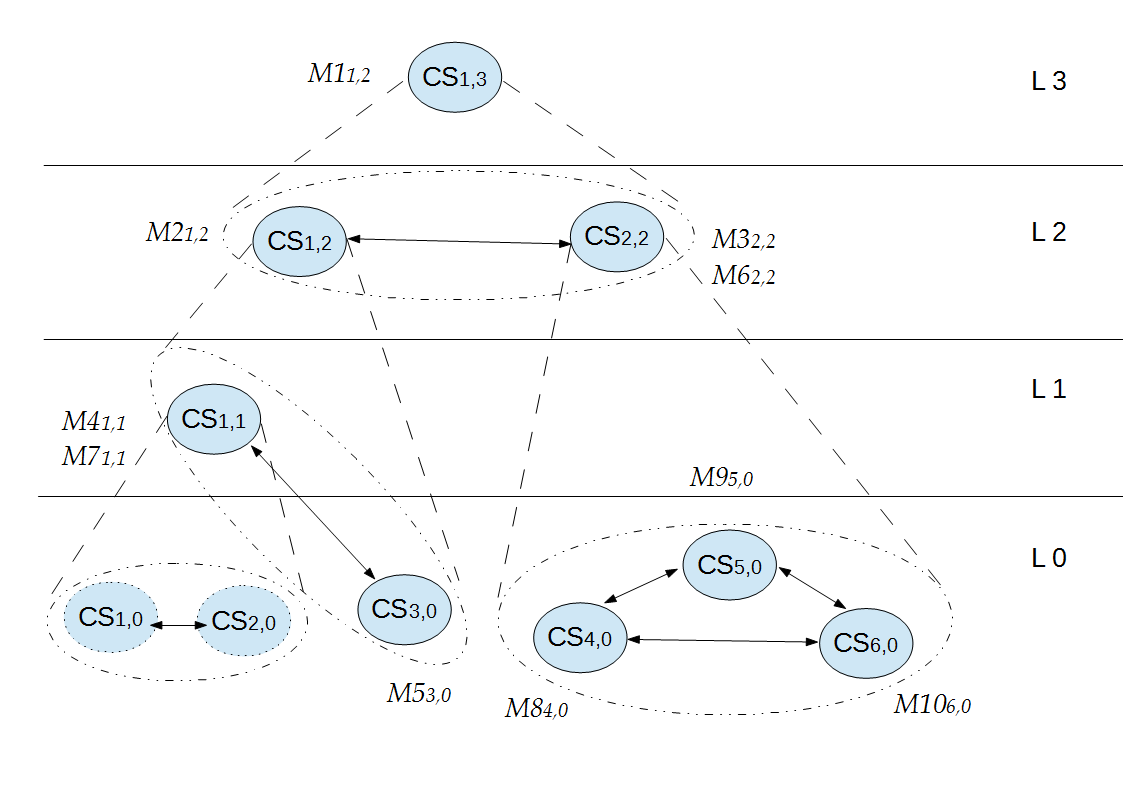}
\caption{Partial allocation of missions over the CS SoS.}
\label{partialDistribution}
\end{figure}

It has to be noted that $CS_{1,0}$ and $CS_{2,0}$ cannot have any allocated mission
because they are not considered as CS but as virtual CS.\\

A functional specification adapted to our formalism can be defined as follows:\\

$fs = <\mathcal{G},\mathcal{M},\mathcal{P},mo>$, with

\begin{itemize}
\item{$\mathcal{G}$}: the set of goals of the SoS.

\item{$\mathcal{M}$}: the set of labels for all missions. The missions can be allocated ($\mathcal{M}x_{n,l}$) or not ($\mathcal{M}x$).

\item{$\mathcal{P}$}: the set of global plans resulting from the goal decomposition tree.

\item{$mo : \mathcal{M} \to 2^\mathcal{G}$}: a function specifying for each mission the set of attached goals.\\

\end{itemize}

Once the functional specification is established the SoS instantiates groups corresponding to goal needs in terms of capacity and available agents,
using available group specifications.
When all group are formed and the elementary CS agent matches the lower level groups,
all groups can be instantiated by a CS agent.
Then the SoS is fully formed and can accomplish its global goal.\\

\begin{algorithm}[!ht]
\caption{SoS reconfiguration}
\label{alg1}
\begin{algorithmic}
\REQUIRE the set of levels: $L$, 
the set of CS agent definition: $CS_{def}$, the set of group specifications: $Gr_{spe}$, the functionnal definition of the soS: $fs_{SoS}$,
 the hierarchical graph: $Gr_L$,
the set of all existing CS agent: $CS$ regrouping
the set of elementary CS: $CS_0$ and the top CS: $CS_H$.
\ENSURE the new SoS configuration respects SoS fundamental characteristics \OR there is no available configuration.

\WHILE{$\exists CS_{n,l} \in CS \wedge CS_{n,l} \notin CS_0 \wedge CS_{n,l}(\mathcal{CS}_{n'',l-1}) = \emptyset$}
\STATE Take a $CS_{n,l}$ of the higher possible level $l$.
\STATE Take the goal $g$ of $CS_{n,l}$ using $\mathcal{M}$ and $mo$ of $fs_{SoS}$.
\STATE Decompose $g$ into a set of subgoals $\mathcal{g}$ present in level(s) $\mathcal{L}$ and create the associated missions $mg$.
\STATE Place $\mathcal{g}$ in $\mathcal{G}$ of $fs_{SoS}$, place $mg$ in $\mathcal{M}$ and $mo$ of $fs_{SoS}$.
\STATE Update $\mathcal{P}$ of $fs_{SoS}$ with the new goal decomposition.
\STATE Select a group specification $gs \in Gr_{spe}$ which can represent $CS_{n,l}$.
\STATE Select CS agent definitions in $CS_{def}$ which can play the necessary roles in $gs$ and can fulfil $\mathcal{g}$.
\STATE Instantiate previous CS agent definitions and allocate $mg$ to them.
\ENDWHILE

\RETURN $fs_{SoS}$
\end{algorithmic}
\end{algorithm}

\subsubsection{Capacity Evolution}

As said previously CS agent capacities evolve through time.
Sometimes CS can lost capacities which compromise their mission and possibly the mission of their super CS.
When the capacities of a CS are modified, its super CS capacities are calculated thanks to CS and group specification.
The capacity evolution is spread through levels thanks to influences.
That means all CS agents with under CS can calculate its own capacity from its under CS capacities and group specification.
So when a CS loses its capacities, it first tries to reorganize itself to stay able to fufil its allocated mission and
if it is impossible, it informs its super CS which do as exactly the same until the top CS agent is informed.\\

If the top CS agent is not able to reorganize itself anymore, by adopting a new organization structure,
it is able to inform the modeler that the SoS has failed and there is no available mean to reach the global goal.\\

\begin{figure}[!ht]
\centering
\includegraphics[width=1\columnwidth]{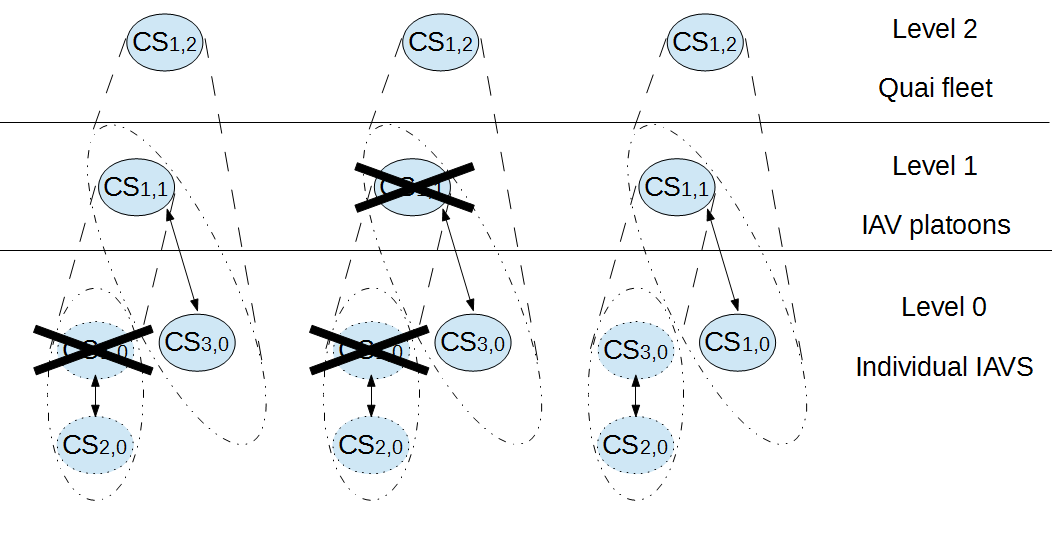}
\caption{Reorganization of a SoS guided by the CS capacity evolution.}
\label{sosReorganization}
\end{figure}

This mechanism can be illustrated by the following example.
The IAV $CS_{1,0}$ has a failure and its capacity to carry a 40-feet container is compromised.
So the capacity of the IAV platoon $CS_{1,1}$ is also modified and
it cannot reorganize itself only with its under CS $CS_{1,0}$ and $CS_{2,0}$.
It can no more accomplish its mission to deliver a series of 40-feet containers to the boat.
So the failure is spread to $CS_{1,2}$, the IAV fleet in quay 1.
$CS_{1,2}$ can reorganize itself by switching IAVs $CS_{3,0}$ and $CS_{1,0}$.
That what is done, maintaining the structure of $CS_{1,2}$.\\

\subsubsection{Particular case of virtual CS instanciation}

In this section we consider the particular case of the creation of a non elementary CS whose under CS lost their CS status.\\

In a MAS model representing SoS, every CS has to be instantiated by a CS agent which respects the above CS definition and its characteristics.
In particular, the identified elementary CS and the top CS have to possess their own instantiating CS agent during all the modelisation.
When a CS stops to respect the SoS characteristics, it ceases to be a CS but it and its under CS interact with the SoS,
even if, they can be no more considered as CS.
That is why agents representing these CS should be deleted in the model but replaced with virtual CS agents which indicates their status of potential CS.\\

\begin{figure}[!ht]
\centering
\includegraphics[width=1\columnwidth]{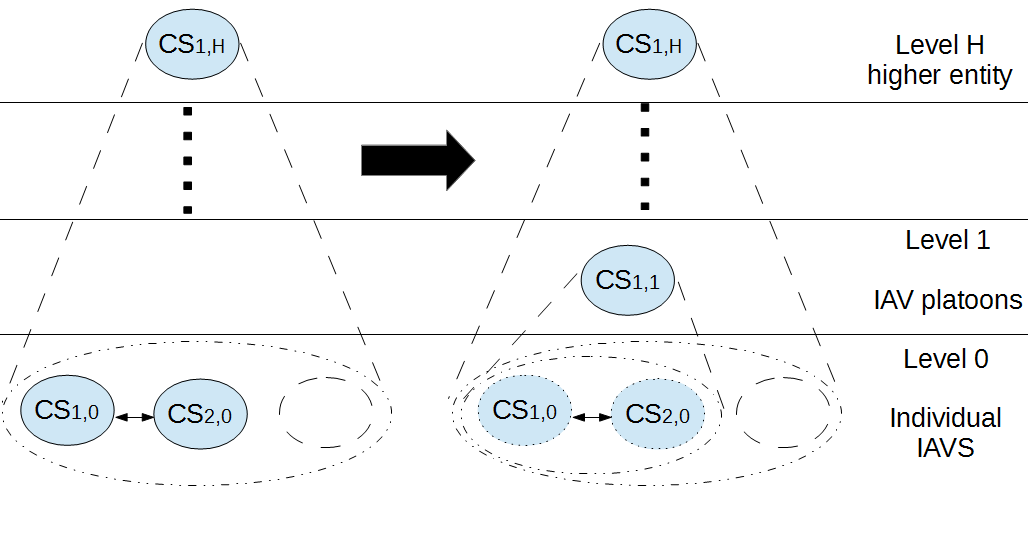}
\caption{Creation of super CS by aggregating under CS in group(s) and updating their status.}
\label{virtualCS}
\end{figure}

To illustrate this, let take a simple example.
Two CS agents, $CS_{1,0}$ and $CS_{2,0}$, representing individual IAVs as elementary CS in a level $l_0$ and another agent,
$CS_{1,1}$, representing the platoon formed by these two vehicles as a super CS in level $l_1$, higher than $l_0$.
When the two AIVs are linked together to carry a heavy container, and one of them became the leader of this tasks,
$CS_{1,0}$ and $CS_{2,0}$ lost their status of CS because they are violating operational and
managerial independences and geographical dispersion characteristics.
$CS_{1,1}$, after the aggregation of $CS_{1,0}$ and $CS_{2,0}$,  represents an elementary CS because its under CS have lost their CS status.
Thus CS agents $CS_{1,0}$ and $CS_{2,0}$ are replaced by virtual CS agents because they are still used to determine $CS_{1,1}$ evolution and
 capacities and can represent again non virtual CS in the future, when the aggregation link between $CS_{1,0}$ and $CS_{2,0}$ will be deleted.

\subsection{Illustration of Dynamic Aspects}

To illustrate the dynamic aspects we consider once again the SoS presented in section II and
whose mission distribution was exposed in fig. \ref{partialDistribution} with our formalism.\\

Before the SoS is built its model should only contain the top CS at the highest level, $CS_{1,3}$ and all 
its elementary CS at the lowest level, $CS_{1,0}$, $CS_{2,0}$, $CS_{3,0}$, $CS_{4,0}$, $CS_{5,0}$ and $CS_{6,0}$.
Other elements of the model should be a set of levels, with hierarchical, influence and perception graphs to aprehend links between levels,
a set of CS agent definitions for every levels, a set of group specifications
corresponding to these agents definition and a nearly empty functional specification.
Only the global goal of the SoS should be known and allocated to the top CS.
And its functional specification should be:

\begin{eqnarray*}
fs_{SoS} = \langle \{g1_3\}, 
\{\mathcal{M}1_{1,3}\},
\emptyset,
\{\mathcal{M}1 \to \{g1_3\}\}
\rangle
\end{eqnarray*}

According to fig. \ref{goalDecomposition} and \ref{partialDistribution}, once the SoS model is fully built, its functional specication should be:

\begin{eqnarray*}
fs_{SoS} = \langle \{g1_3,g2_2,g3_2,g4_1,g5_0,g6_2,g7_1,g8_0,\dots\},  \\
\{\mathcal{M}1_{1,3},\mathcal{M}2_{1,2},\mathcal{M}3_{2,2},\mathcal{M}4_{1,1},\mathcal{M}5_{3,0},
\mathcal{M}6_{2,2},\mathcal{M}7_{1,1},\\ \mathcal{M}8_{4,0},
\mathcal{M}9_{5,0},\mathcal{M}10_{6,0}\dots\},
\{"g1_3=g2_2,g3_2",\\ "g2_2=g4_1,g5_0",
"g3_2=g6_2",\dots\},
\{\mathcal{M}1 \to \{g1_3\},\\ \mathcal{M}2_{1,2} \to \{g2_2\},
\mathcal{M}3_{2,2} \to \{g3_2\}, \mathcal{M}4_{1,1} \to \{g4_1\},\\
 \mathcal{M}5_{3,0} \to \{g5_0\}, \dots, \mathcal{M}7_{1,1} \to \{g7_1,g8_1,g9_1,g10_1\},
\dots \} \rangle
\end{eqnarray*}

\subsection{Respecting Fundamental characteristics}

In this section we show that directed SoS characteristics are respected in agent-based models generated using our formalism.\\

It is important to consider that, in the first three characteristics, the CS agents $CS_{i',j'}$ and $CS_{i'',j''}$
of a system $CS_{i,j}$ are not related by a direct hierachical relation:\\

\begin{eqnarray*}
\forall CS_{i',j'}, CS_{i'',j''} \in  CS_{i,j},\\
CS_{i',j'} \notin CS_{i'',j''} \wedge 
CS_{i'',j''} \notin CS_{i',j'} \\
\end{eqnarray*}

Here are the five characteristics that any SoS should respect (see section I) and their translation in a MAS model representing a SoS:\\

\begin{enumerate}

\item{\textbf{Operational independence}}:  CS agents $CS_{i',j'}$ and $CS_{i'',j''}$ of a system $CS_{i,j}$ are independent in an operational way iff
they are instanciated by CS agents, $\mathcal{A}$, and
possess their own private variables, $s_a$, representing their resources.

\begin{eqnarray*}
\forall CS_{i',j'}, CS_{i'',j''} \in  CS_{i,j},\\
CS_{i',j'}, CS_{i'',j''} \in \mathcal{A} \wedge \\
s_{CS_{i',j'}} \neq \emptyset \wedge s_{CS_{i'',j''}} \neq \emptyset \wedge \\
s_{CS_{i',j'}} \cap s_{CS_{i'',j''}} = \emptyset
\end{eqnarray*}


Fig. \ref{graphicalRepresentationSimple2} shows that $CS_{4,0}$ is operationally independent,
but $CS_{1,0}$ is not, because it shares a part or its resources with $CS_{2,0}$.\\

\item{\textbf{Managerial independence}}: CS agents $CS_{i',j'}$ and $CS_{i'',j''}$ of a system $CS_{i,j}$ are independent in a managerial way iff
they do not share any part of their own missions, $\mathcal{M}_{i',j'}$ and $\mathcal{M}_{i'',j''}$.

\begin{eqnarray*}
\forall CS_{i',j'}, CS_{i'',j''} \in  CS_{i,j},\\
\mathcal{M}_{i',j'} \cap \mathcal{M}_{i'',j''}  = \emptyset
\end{eqnarray*}

In non specific SoS, this characteristic says that for each CS agent, no other CS agent gives it direct orders and only its super CS agent can allocate missions to it.
But because we consider directed SoS endowed with a strict hierarchy, the above notation is sufficient.
Each CS agent manages its own missions independently of other CS agents.\\

$CS_{4,0}$ of fig. \ref{graphicalRepresentationSimple2} is independent in an managerial way,
but $CS_{1,0}$ is not, because it shares its missions with $CS_{2,0}$.\\

\item{\textbf{Geographic distribution}}: CS agents $CS_{i',j'}$ and $CS_{i'',j''}$ of a system $CS_{i,j}$
are geographically distributed iff
their physical state, $\phi_a$, are totally distinct and not directly dependent.
This last property can be formulated by the fact that it is not possible
to compute directly all or any part of the physical state of a CS agent (using a function $f$)
from all or any part of the physical state of another CS agent.

\begin{eqnarray*}
\forall CS_{i',j'}, CS_{i'',j''}  \in CS_{i,j},\\
\phi_{CS_{i',j'}} \cap \phi_{CS_{i'',j''}} = \emptyset \wedge\\
\big( \forall \phi'_{CS_{i',j'}} \subset \phi_{CS_{i',j'}},
\forall \phi'_{CS_{i'',j''}} \subset \phi_{CS_{i'',j''}},\\
\nexists f:\Phi \to \Phi, f(\phi'_{CS_{i',j'}}) = \phi'_{CS_{i'',j''}}\\
 \vee f(\phi'_{CS_{i'',j''}}) = \phi'_{CS_{i',j'}}\big)
\end{eqnarray*}

In this case, no physical link between CS agents is authorized to accomplish missions.
Only information exchanges between CS agents are authorized.\\

Practically, in a agent-based model it also can be noticed that if each CS agent is situated in an environment
endowed with a metric and is not physically linked to other CS agents, according to this metric.\\

Fig. \ref{graphicalRepresentationSimple2} shows that $CS_{4,0}$ is geographically dispersed,
but $CS_{1,0}$ and $CS_{2,0}$ are not.
Because, in the case of $CS_{1,0}$ is the leader of the platoon $CS_{1,1}$
and $CS_{2,0}$ is a follower, then it exists a function $f$ and $f(pos_{CS_{1,0}}) = pos_{CS_{2,0}}$,
where $pos_{CS_{1,0}}$ and $pos_{CS_{2,0}}$ designate the part of the physical state of these two CS agents
which represents their position.\\ 

\begin{figure}[!ht]
\centering
\includegraphics[width=1\columnwidth]{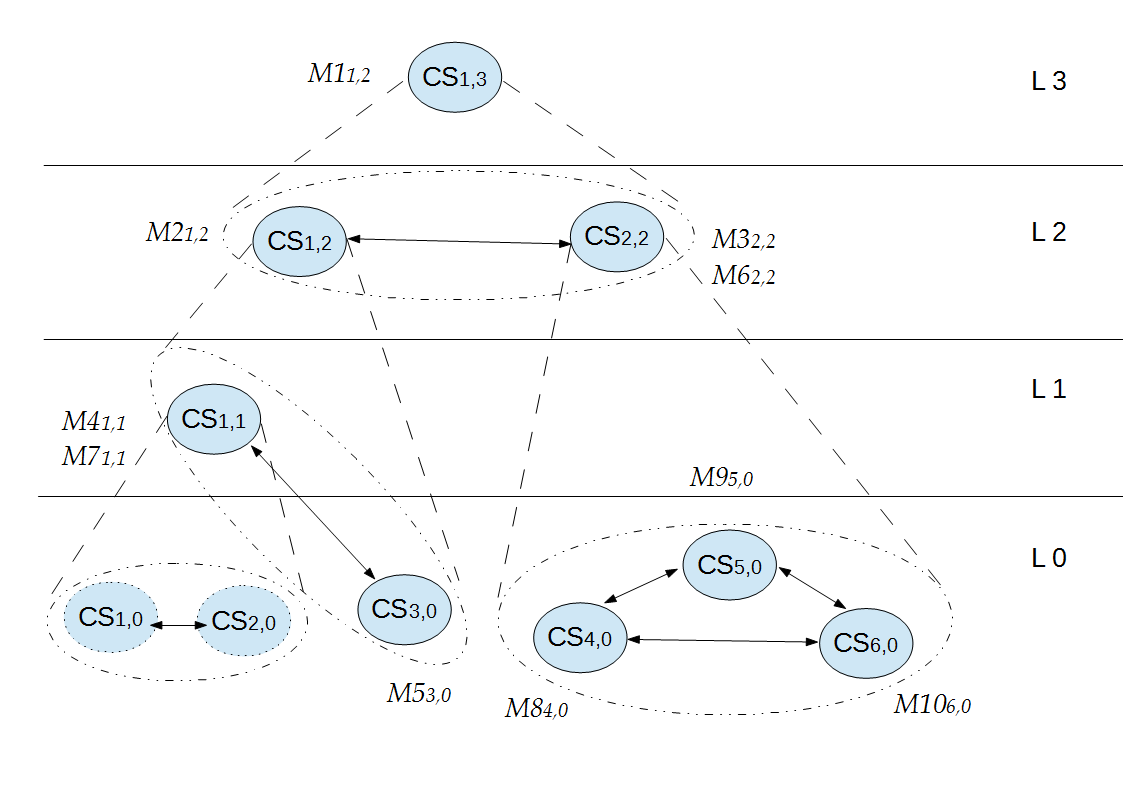}
\caption{Characteristics of a SoS.}
\label{graphicalRepresentationSimple2}
\end{figure}

\item{\textbf{Emergent Behaviour}}: A set of CS agents $CS_{i',j'}$ of a system $CS_{i,j}$  cooperates iff 
CS agents cooperate to realize a global mission $\mathcal{M}_{i,j}$ of the system formed of a set of missions $\mathcal{M}_{i',j'}$,
that each CS agent of $CS_{i,j}$ cannot realize alone.\\

\begin{eqnarray*}
\forall CS_{i',j'}  \in CS_{i,j},\\
\bigcup {\mathcal{M}_{i',j'}} = \mathcal{M}_{i,j}
\end{eqnarray*}

Said in another way, an agent based model represents a SoS iff joint actions of several CS agents is obligatory necessary to reach the global goal.\\

CS agent $CS_{2,2}$ of fig. \ref{graphicalRepresentationSimple2} cannot fulfil its missions $\mathcal{M}3_{2,2}$ and $\mathcal{M}6_{2,2}$ only through
the cooperation of its under CS $CS_{4,0}$, $CS_{5,0}$ and $CS_{6,0}$ wich fulfil missions $\mathcal{M}8_{4,0}$, $\mathcal{M}9_{5,0}$ and $\mathcal{M}10_{6,0}$.\\

\item{\textbf{Evolutionary development}}: 
This characteristic is different from the others because it can be moticed only during the SoS execution.
In particular, this characteristic is expressed in three situations:
a) a CS agent is added to the SoS, b) a CS agent of the SoS is removed or 
c) a CS agent notices that one of its missions is impossible to fulfil according to its capacity function.\\

To allow a SoS to accept new CS agents like in situation a) it is neccessary that the new CS agents are placed in groups
whose specification contains roles which do not affect the capacities of the CS agent formed by that group.
In this case, the addition of new CS agents does not affect the capacity of the SoS to fulfil its global goal.
To take avdantage of this addition, to enhanche the capacity of the SoS, an algorithm is necessary to optimally
reorganize the SoS, including the new CS agent.
This kind of algorithm is very specific to the design of the SoS problem and this not the topic of this article.
Therefore, it is not developed here.\\

Situations b) and c) are quite similar. Because when the deletion of a CS agent doesn't affect its super CS capacity,
it doesn't affect the capacity of the whole SoS. And when it affects its super CS capacity this super CS is in situation c).
In this case the loss of capacity has to be communicated via influences, impossible missions have to bet set free and
the SoS reorganization has to be performed like in previous algorithm.\\

A system $CS_{i,j}$, formed by several CS agents $CS_{i',j'}$ follows an evolutionary development
iff when one of its CS agents allocated mission $\mathcal{M}x_{i',j'}$ is detected as impossible
to fufil according to its capacity function $\delta$ and the atcutal state of the level $j'$ $\lambda_{i,j}$ and $\delta_{j'}(t)$,
this mission is no more allocated to any CS agent,
but is set free in the set of missions $\mathcal{M}$ in the functional specification of the SoS.

\begin{eqnarray*}
\forall CS_{i',j'}  \in CS_{i,j},\\
\lambda_{i,j}(\delta_{j'}(t),\delta(\mathcal{M}x_{i',j'}))= 0\\
\to \mathcal{M}x_{i',j'} \notin \mathcal{M} \cup \mathcal{M}x \in \mathcal{M}
\end{eqnarray*}

Once the mission is set free an algorithm has to allocate it to another CS agent or reorganize the
SoS to fufil it, in a feasible way.
\cite{Khalil:2012a} presents such SoS reorganisations whithout evoking a formalized algorithm.\\ 


 Fig. \ref{evolutionaryDevelopment} shows three steps in the life of a SoS.
 The second step is obtained by the suppression of $CS_{3,0}$.
 Step three represents the same SoS after the addition of $CS_{3,0}$ and $CS_{4,0}$ to the SoS.
 It should be noted that $CS_{3,0}$ from step one and three is not obligatory the same CS agent.
 
\end{enumerate}

\begin{figure}[!ht]
\centering
\includegraphics[width=1\columnwidth]{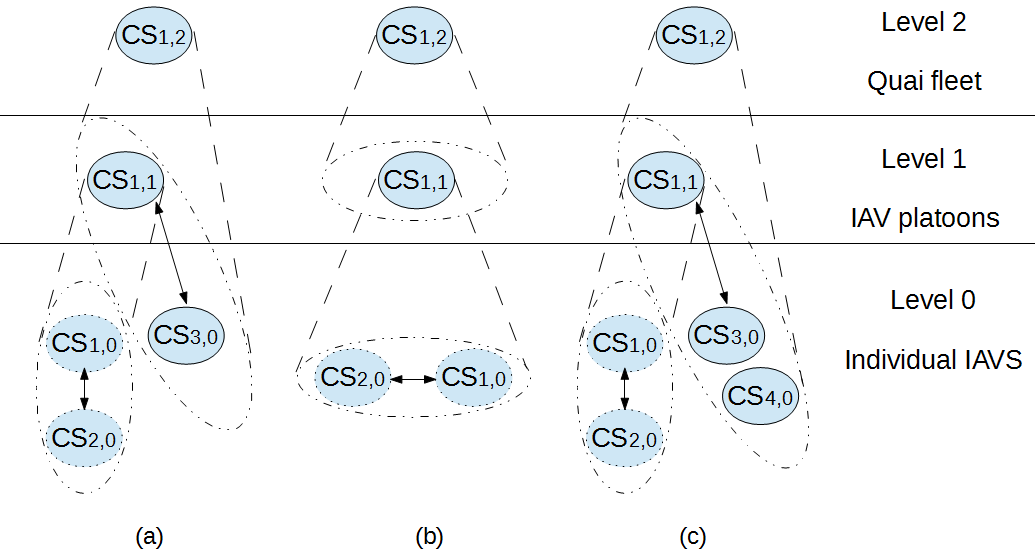}
\caption{Three steps in the life of a SoS: a) initial state, b) deletion of $CS_{3,0}$ and c) addition of $CS_{3,0}$ ans $CS_{4,0}$.}
\label{evolutionaryDevelopment}
\end{figure}

\section{Conclusion and Future Work}

In this article we presented an operational SoS definition based on the fundamental characteristics for CS:
operational independence, managerial independence, geographic distribution, evolutionary development and emergent behaviour.
We proposed a generic multi-level multi-agent formalism to represent SoS managed by a central authority.
The proposed model can be applied to any SoS controling it in a changing environment affecting CS capacities.
Only repeated execution of multi-agent based simulations (MABS) seems suited to test and validate algorithms to control,
diagnose systems with numerous heterogenous interacting sub-systems endowed of a changing hierarchical structure.
Moreover the system is cut into groups and levels.
We also give elements to prove the validity of our approach.\\

\begin{figure}[!ht]
\centering
\includegraphics[width=1\columnwidth]{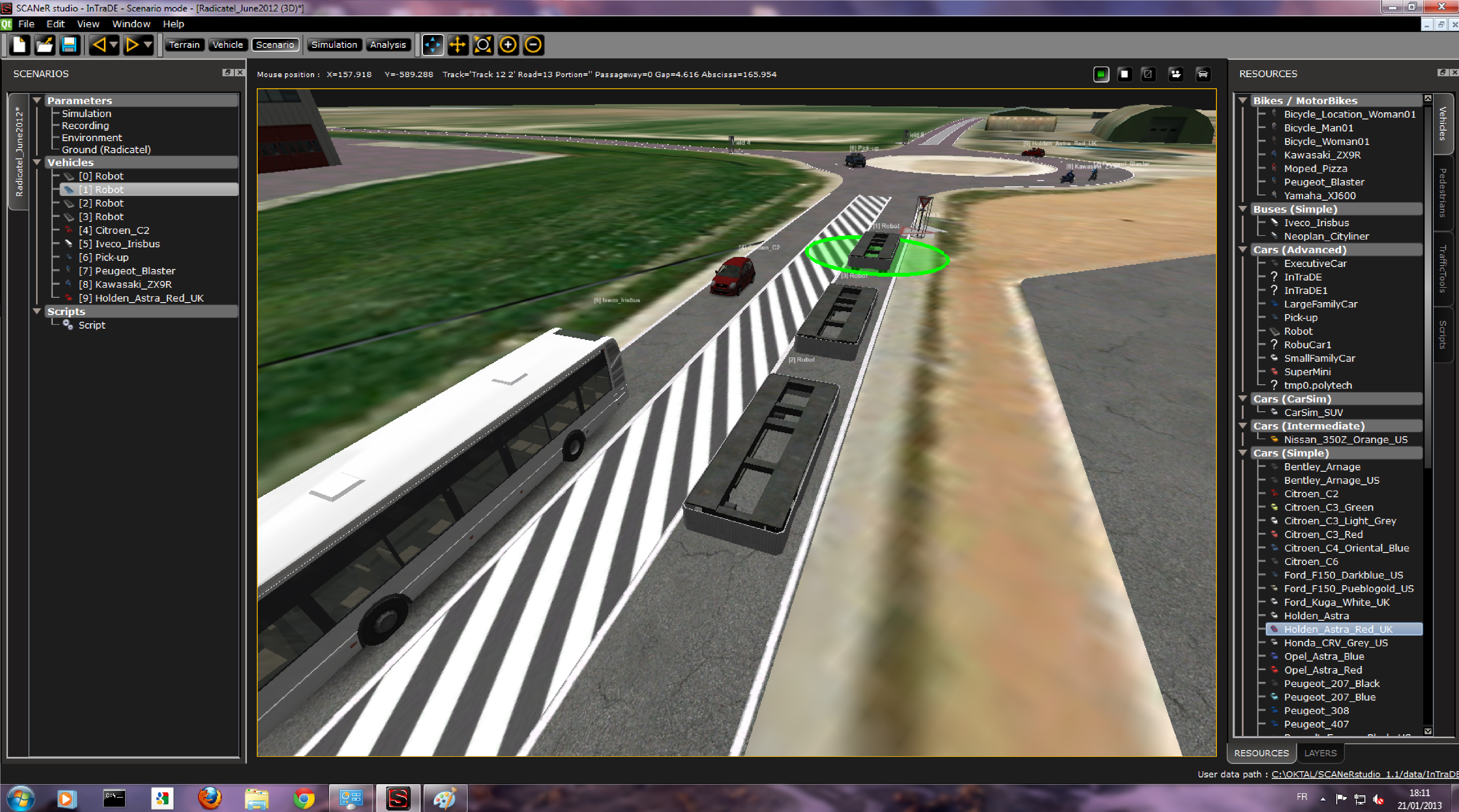}
\caption{A platoon of IAvs in SCANeRstudio.}
\label{intradeView}
\end{figure}

These concepts are applied by modeling SoS whose elementary CS are IAVs in the InTrade project \cite{intrade}.
The resulting simulation is being implemented on MaDKit multi-agent platform endowed with AGR model \cite{madkit}.
SCANeRstudio \cite{scanerstudio} is a simulation software which allows to pilot AIVs in realistic conditions.
The resulting simulation should be use to control direct system by interfacing a SoS model of IAVs fleet in
MaDKit and a realistic representation of IAVs in a container port in SCANeRstudio.
In particular, it permits to observe the influence of low level events managed by SCANeRstudio at the same level or in higher levels.
It could be an IAV failure spreading to its platoon or the complete loss of a vehicule wich force a fleet of IAV or the whole SoS to recalculate their scheduling.
\\

\begin{figure}[!ht]
\centering
\includegraphics[width=1\columnwidth]{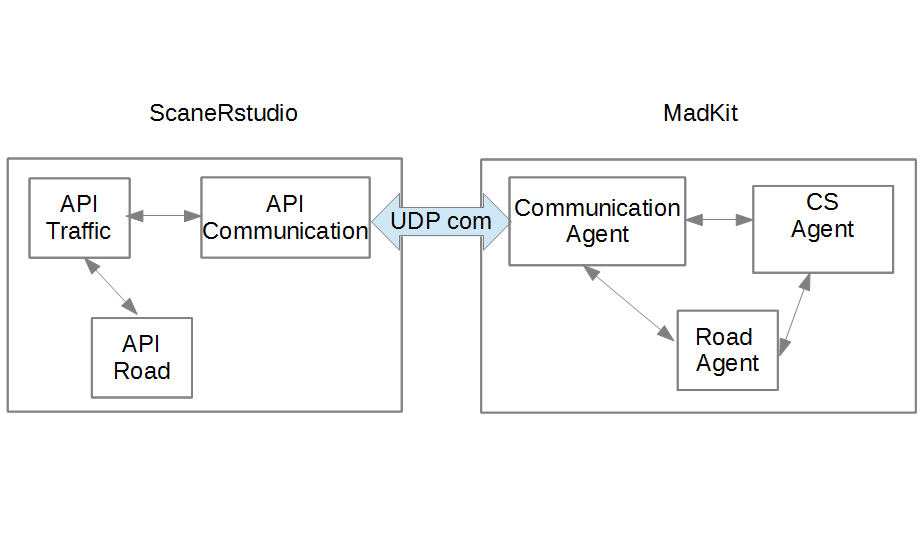}
\caption{Relations between SCANeRstudio and Madkit via UDP communications.
SCANeRstudio is a network of communicating APIs and Madkit is a multi-agent platform.}
\label{application}
\end{figure}

A developpement of the applied part of this work is to couple the multi-level multi-agent based model in MaDKit with other simulation tools than SCANeRstudio.
Flexsim and its Container Terminal Library (Flexsim CT) is used to model traffic flow in container terminal environments \cite{flexsim}.
It could figure the same SoS but represented at a higher level and providing semi-realistic inputs for that level.
For the moment, levels of the IRM4MLS are only considered for simulating different scales of granularities.
But IRM4MLS offers the possibility to model levels in parallel, representing relatively independant aspects of a phenomenon or SoS and simultaneously activated.
This can be experimented using InTrade examples, creating levels reserved for diagnostic and other driving of IAVs.
A SoS of IAVs considered through these two functions will lead to two different structures.  
In the future, our work could be applied to other forms of SoS but not only directed ones with addition of authority and organizational aspects.
Another track is to improve the building group mechanism.
It can be done by specifying strategies to optimize the group specification choice and/or the choice of CS agents
to constitute these groups according to the robustness of the resulting SoS or the quality of the accomplished mission.

\ifCLASSOPTIONcaptionsoff
  \newpage
\fi



\bibliographystyle{IEEEtran}
\bibliography{maBiblio} 

%








\end{document}